\documentclass[12pt]{article}

\usepackage{graphicx}
\usepackage{epsfig}

\begin{document}
\begin{center}
{\large\bf SELECTION TRIGGER FOR RARE}\\

{\large \bf QUARK-GLUON PLASMA FORMATION EVENTS}

\vspace{0.8cm}

{\large\bf G.H. Arakelyan$^1$, C. Merino$^2$, and Yu.M. Shabelski$^{3}$}\\

\vspace{0.5cm}
\bf$^1$A.Alikhanyan National Scientific Laboratory\\
Yerevan Physics Institute\\
Yerevan, 0036, Armenia\\
E-mail: argev@mail.yerphi.am\\

$^2$Departamento de F\'\i sica de Part\'\i culas, Facultade de F\'\i sica\\
and Instituto Galego de F\'\i sica de Altas Enerx\'\i as (IGFAE)\\
Universidade de Santiago de Compostela\\
Santiago de Compostela 15782\\
Galiza, Spain\\
E-mail: merino@fpaxp1.usc.es \\

\vspace{.2cm}

$^{3}$Petersburg Nuclear Physics Institute\\
NCR Kurchatov Institute\\
Gatchina, St.Petersburg 188350, Russia\\
E-mail: shabelsk@thd.pnpi.spb.ru
\vskip 0.9 truecm

\vspace{1.2cm}

\bf A b s t r a c t
\end{center}

We consider the experimental ratios of multistrange to strange antibaryon
production and compare them to the standard Quark-Gluon String Model
predictions. The significant differences between the experimental
$\overline{\Xi}^+/\overline{\Lambda}$ and, especially,
$\overline{\Omega}^+/\overline{\Lambda}$ values and model predictions can be 
interpreted as the signal of the existence of an additional source of
multistrange antibaryon production. The possible connection of this
additional source with Quark-Gluon Plasma (QGP) formation can turn  
the $\overline{\Omega}^+$ production into a good QGP signature.      

\vskip 1.5cm


\newpage

\section{Introduction}

The Quark-Gluon String Model (QGSM) \cite{KTM,KaPi} is based on the Dual
Topological Unitarization (DTU), Regge phenomenology and nonperturbative
notions of QCD. This model is successfully used for the description of
multiparticle production processes in hadron-hadron
\cite{KaPi,Sh,AMPS,MPS},
hadron-nucleus \cite{KTMS,Sh1}, and nucleus-nucleus \cite{Sha,Shab,JDDS}
collisions. In particular, the rapidity dependence of the inclusive densities
of different secondaries ($\pi^{\pm}$, $K^{\pm}$, p, and $\overline{p}$) \cite{JDDS}
produced in $Pb+Pb$ collisions, and of the net 
baryon ($p-\overline{p}$ and $\Lambda - \overline{\Lambda}$) \cite{???} on nuclear 
targets at CERN SpS energies, have been reasonably described in the framework
of QGSM. 
 
In QGSM high energy interactions are considered as proceeding via
the exchange of one or several Pomerons. The cut of the elastic scattering
amplitude determines the particle production processes which occur via 
production and subsequent decay of the quark-gluon strings.

In the case of interaction with a nuclear target, the Multiple Scattering
Theory (Gribov-Glauber Theory) is used.
For nucleus-nucleus collisions, the Multiple Scattering Theory
also allows to consider this interactions as the superposition of separate
nucleon-nucleon interactions. However, in this case the analytical
summation of all the diagrams is impossible~\cite{BSh}. The significant
classes of diagrams can be summed up analytically in the so-called rigid
target approximation \cite{Alk}, which will be used in the present paper.

At very high energies, the contribution of enhancement Reggeon diagrams 
(percolation effects) becomes important, what leads to a new phenomenological
effect, the suppression of the inclusive density of secondaries \cite{CKTr}
into the average central (midrapidity) region. This corresponds to a significant
fusion of the produced quark-gluon strings. In the energy limit when the
probability of this fusions is very large, one should expect the appearance
of a new state of matter, the Quark-Gluon Plasma(QGP). However, the process of
QGP production can already occur with small probability at a not so high
energy. 

In this paper we present the ratios of multistrange to strange antihyperon 
production in the collision of projectil A (nucleon or nucleus) with a nuclear 
target B. Let us define: 
\begin{equation}
R(\overline{\Xi}^+/\overline{\Lambda}) = \frac{dn}{dy}(A+B\to\overline{\Xi}^++X)/ \frac{dn}{dy}
(A+B\to\overline{\Lambda}+X) \;,
\end{equation} 
\begin{equation}
R(\overline{\Omega}^+/\overline{\Lambda}) = \frac{dn}{dy}
(A+B\to\overline{\Omega}^++X)/ \frac{dn}{dy}
(A+B\to\overline{\Lambda}+X) \;.
\end{equation} 
The produced antihyperons, $\overline{\Xi}^+$ and $\overline{\Omega}^+$, contain  
valence antiquarks newly produced during the collision.
The ratios in Eqs. (1) and (2) are reasonably described by QGSM in the cases
when a not very large number of incident nucleons participate in the collision
(nucleon-nucleus or peripheral nucleus-nucleus collisions).

The number of quark-gluon strings (cut pomerons) in nucleus-nucleus
collisions increases with centrality. If the secondaries are produced independently 
in every quark-gluon string, the ratio of yields of different 
particles should not depend on centrality. However, experimentally the enhancement
of the yield of $\overline{\Xi}^+$ is stronger than that of $\overline{\Lambda}$,
and for $\Omega^- + \overline{\Omega}^+$ the enhancement is also stronger than for
$\overline{\Xi}^+$ (e.g. see \cite{HUA}). This means that an additional source of 
multistrange hyperons appears from collective interactions of several strings,
and this additional source can be considered as a QGP state effect.

In the present paper we consider the central Pb+Pb collisions. The
experimental ratios of Eqs. (1) and (2), as functions of centralities,
are in total disagreement with the standard QGSM predictions, where
it is assumed that each secondary particle is produced by the independent 
fragmentation of every quark-gluon string.
We can now assume that some new contribution to the multistrange antihyperon 
formation appears in this case of central Pb+Pb collisions (see Sections 3
and 4). 
This new contribution, that is not accounted for in the standard 
QGSM scheme, could be seen as a signature of QGP production in some small
volume of the interaction region.

\section{Production of Secondaries in the QGSM}

The QGSM \cite{KTM,KaPi} allows one to make
quantitative predictions of different features of multiparticle production,
in particular, the  inclusive densities of different secondaries both in
the central and in the beam fragmentation regions. In QGSM ,high energy
hadron-nucleon collisions are implemented through the exchange
of one or several Pomerons, all elastic and inelastic processes resulting
from cutting through or between Pomerons~\cite{AGK}.

Each Pomeron corresponds to a cylindrical diagram, and thus, when cutting
one Pomeron, two showers of secondaries are produced (see Fig.~1a,1b).
\begin{figure}[htb]
\centering
\vskip -1.cm
\includegraphics[width=.4\hsize]{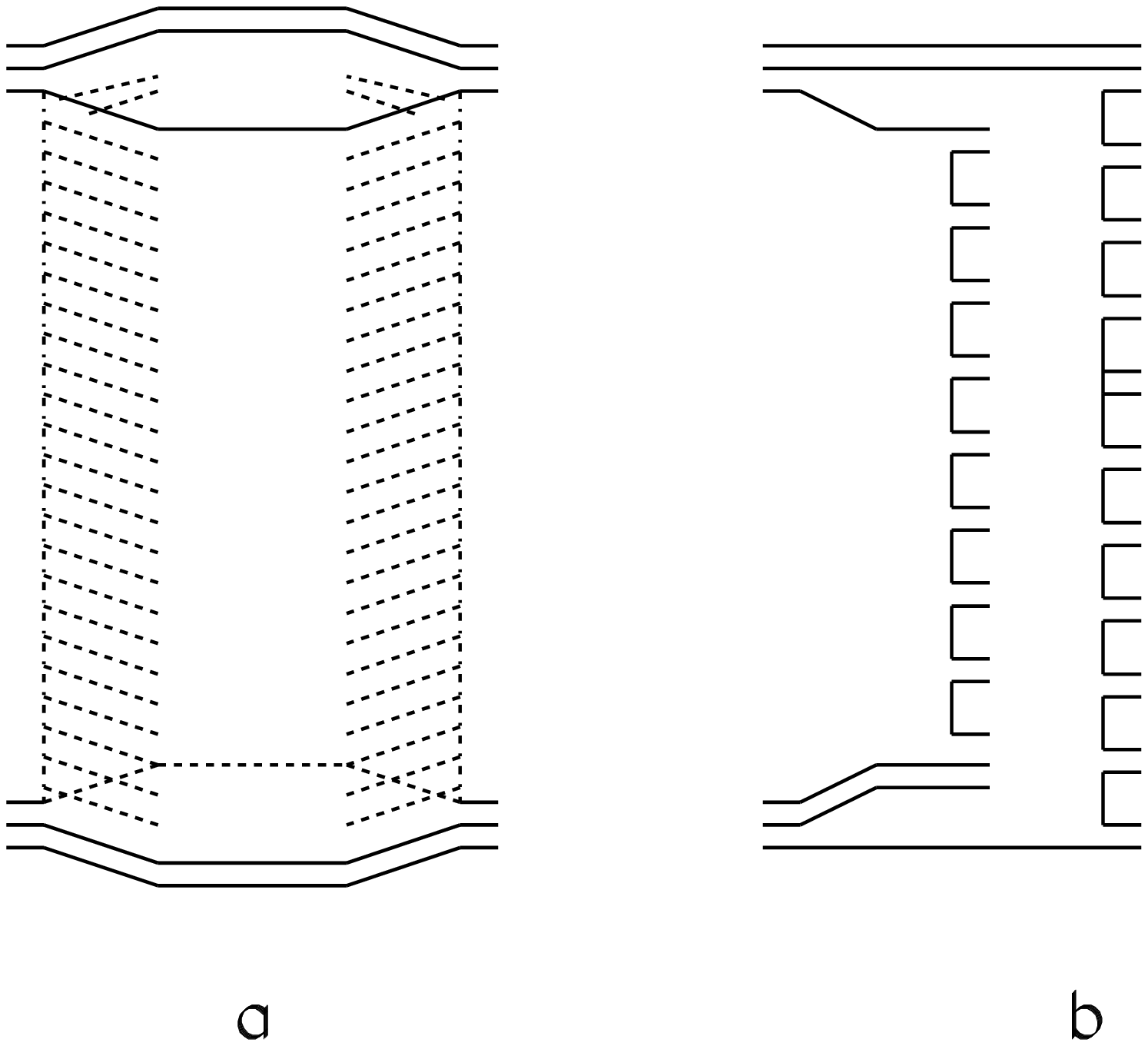}
\vskip -1.7cm
\includegraphics[width=.45\hsize]{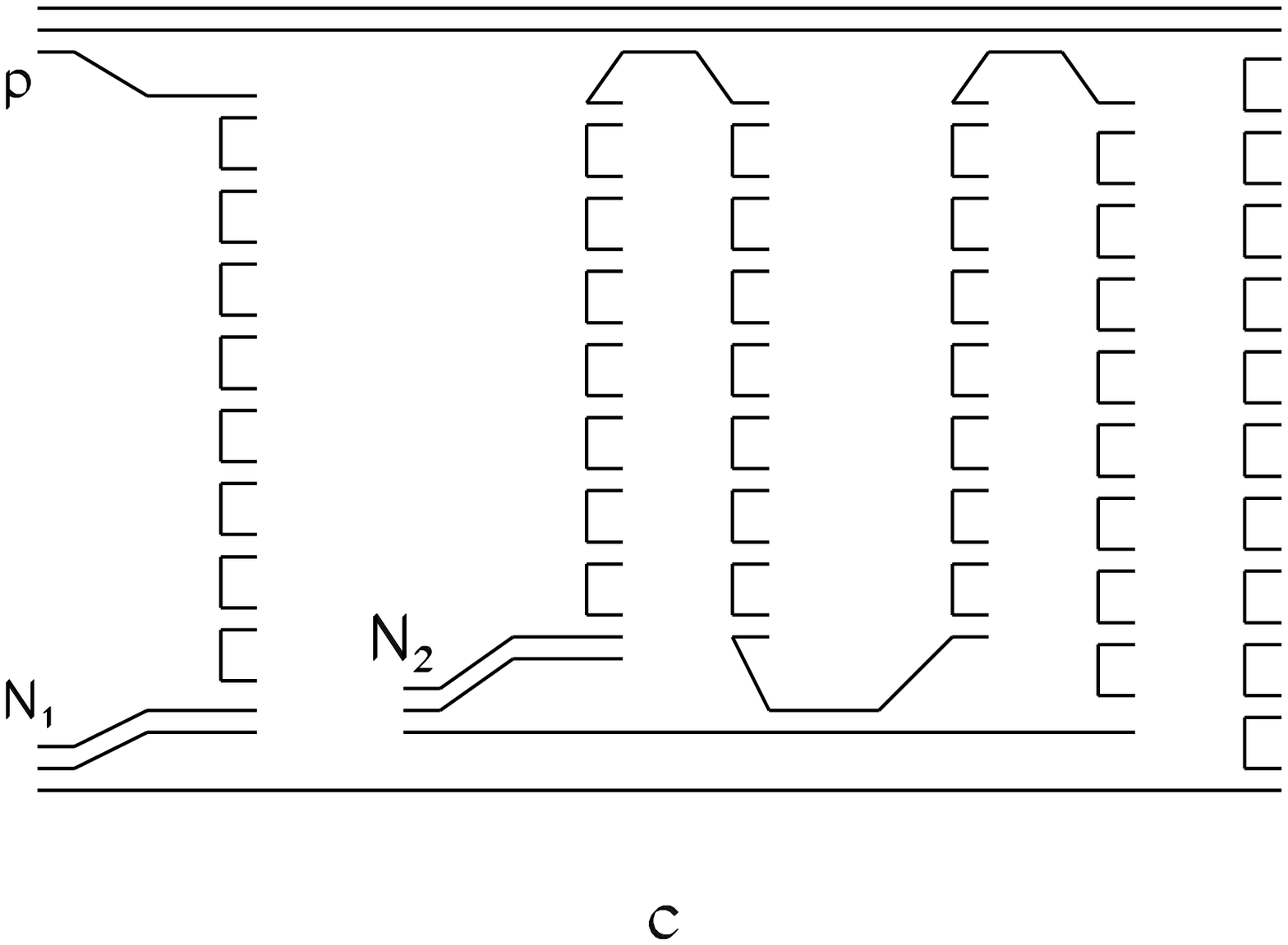}
\vskip -.8cm
\caption{\footnotesize
(a) Cylindrical diagram representing the Pomeron exchange within the DTU
classification. Quarks are shown by solid lines; (b) Cut of the cylindrical
diagram corresponding to the single-Pomeron exchange contribution in inelastic
$pp$ scattering; (c) Diagram corresponding to the inelastic interaction of an
incident proton with two target nucleons $N_1$ and $N_2$ in a $pA$ collision.}
\end{figure}

The inclusive spectrum
of a secondary hadron $h$ is then determined by the convolution of the
diquark, valence quark, and sea quark distributions $u(x,n)$ (where every 
distribution $u_i(x,n)$ is normalized to unity) in the incident
particles, with the fragmentation functions $G^h(z)$ of quarks and diquarks
into the secondary hadron $h$. These distributions, as well as the
fragmentation functions, are constructed using the Reggeon counting rules 
\cite{Kai}.

In particular, in the case of $n > 1$, i.e. in the case of multipomeron
exchange, the distributions of valence quarks and diquarks are softened due
to the appearence of a new contribution from sea quarks-antiquarks. 

The details of the model are presented in \cite{KTM,KaPi,Sh,ACKS}. The
average number of exchanged Pomerons $\langle n \rangle_{pp}$ slowly
increase with the energy. The values of the Pomeron parameters have been
taken from~\cite{Sh}.

For a nucleon target, the inclusive rapidity, y, or Feynman-$x$, $x_F$,
spectrum of a secondary hadron $h$ has the form~\cite{KTM}:
\begin{equation}
\frac{dn}{dy}\ = \
\frac{x_E}{\sigma_{inel}}\cdot \frac{d\sigma}{dx_F}\ = \
\sum_{n=1}^\infty w_n\cdot\phi_n^h (x) + w_D \cdot\phi_D^h (x) \ ,
\end{equation}
where the functions $\phi_{n}^{h}(x)$ determine the contribution of diagrams
with $n$ cut Pomerons, $w_n$ is the relative weight of this diagram, and the
term $w_D \cdot\phi_D^h (x)$ accounts for the contribution of diffraction
dissociation processes.

For $pp$ collisions
\begin{equation}
\phi_n^{h}(x) = f_{qq}^{h}(x_{+},n) \cdot f_{q}^{h}(x_{-},n) +
f_{q}^{h}(x_{+},n) \cdot f_{qq}^{h}(x_{-},n) +
2(n-1)f_{s}^{h}(x_{+},n) \cdot f_{s}^{h}(x_{-},n)\ \  ,
\end{equation}

\begin{equation}
x_{\pm} = \frac{1}{2}[\sqrt{4m_{T}^{2}/s+x^{2}}\pm{x}]\ \ ,
\end{equation}
where $f_{qq}$, $f_{q}$, and $f_{s}$ correspond to the contributions
of diquarks, valence quarks, and sea quarks, respectively.

These contributions are determined by the convolution of the diquark and
quark distributions with the fragmentation functions, e.g.,
\begin{equation}
f_{q}^{h}(x_{+},n) = \int_{x_{+}}^{1}
u_{q}(x_{1},n)\cdot G_{q}^{h}(x_{+}/x_{1}) dx_{1}\ \ .
\end{equation}

In the calculation of the inclusive spectra of secondaries produced in
$pA$ collisions we should consider the possibility of one or several Pomeron
cuts in each of the $\nu$ blobs of proton-nucleon inelastic interactions.
For example, in Fig.~1c it is shown one of the diagrams contributing to the
inelastic interaction of a beam proton with two target nucleons. In the
blob of the proton-nucleon1 interaction one Pomeron is cut, and
in the blob of the proton-nucleon2 interaction two Pomerons are cut. 
The details can be found in \cite{KTMS}.

It is essential to take into account all digrams with every possible Pomeron
configuration and its permutations. The diquark and quark distributions and
the fragmentation functions here are the same as in the case of $pN$
interaction.

The total number of exchanged Pomerons becomes as large as
\begin{equation}
\langle n \rangle_{pA} \sim
\langle \nu \rangle_{pA} \cdot \langle n \rangle_{pN} \;,
\end{equation}
where $\langle \nu \rangle_{pA}$ is the average number of inelastic
collisions inside the nucleus
(about 4 for heavy nuclei at SpS energies).

The process shown in Fig.~1c satisfies~\cite{Sh3,BT,Weis,Jar} the condition
that the absorptive parts of the hadron-nucleus amplitude are determined by
the combination of the absorptive parts of the hadron-nucleon amplitudes.

In the case of a nucleus-nucleus collision, we use in the fragmentation region of
the projectile the approach in refs.~\cite{Sha,Shab,JDDS}, where the beam of
independent nucleons of the projectile interacts with the target nucleus,
what corresponds to the rigid target approximation \cite{Alk} of
Glauber Theory. Correspondingly, in the target fragmentation region the
beam of independent target nucleons interacts with the projectile nucleus.
The results obtained on the two fragmentation regions coincide in the
central region. The corrections for energy conservation play here a very important
role when the initial energy is not very high. This approach was used in~\cite{JDDS}
for the succsessful description  of $\pi^{\pm}$, $K^{\pm}$, $p$, and $\overline{p}$
produced in Pb+Pb collisions at 158 GeV per nucleon.

\section{\bf Inclusive Spectra with String Fusion and 
Percolation Effects}

The QGSM gives a reasonable description \cite{MPS,KTMS,JDDS} of the 
inclusive spectra on nuclear targets at energies $\sqrt{s_{NN}}$ = 14$-$30 GeV.
At RHIC energies the situation drastically changes. The spectra of
secondaries produced in $pp$ collisions are described by QGSM rather well 
\cite{MPS}, but the RHIC experimental data for Au+Au collisions 
\cite{Phob,Phen} give clear evidence of the supression 
effects which reduce the midrapidity inclusive density by about a factor two
when compared to the predictions based on the superposition picture 
\cite{CMT,Sh6,AP}. This reduction can be explained by the inelastic screening
corrections connected to multipomeron interactions~\cite{CKTr} (see Fig.~2).

\begin{figure}[htb]
\centering
\vskip -1.5cm
\includegraphics[width=.42\hsize]{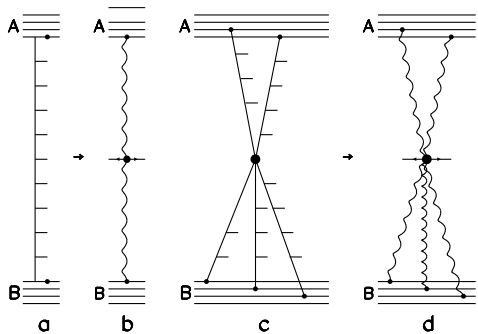}
\caption{\footnotesize
(a) Multipheral ladder diagram; (b) Inclusive cross section to which diagram (a) corresponds;
(c) Diagram with fusion of several ladders; (d) Inclusive cross section to which
diagram (c) corresponds.}
\end{figure}

At fixed target energies, $\sqrt{s_{NN}} \leq $ 30 GeV, the 
nucleus-nucleus interaction can be described as the sum of one-Pomeron and 
of multipomeron eikonal exchanges. The inelastic processes are then determined 
by the 
production of one (Fig.~2a), or several  multiperipheral ladders, and the 
corresponding inclusive cross sections are described by the diagram of Fig.~2b. 

In accordance with the Parton Model~\cite{Kan,NNN}, the fusion of 
multiperipheral 
ladders shown in Fig.~2c becomes essential when the energy increases, what
should reduce the inclusive density of secondaries. Such processes correspond
to the enhancement Reggeon diagrams shown in Fig.~2d, and whith more complicated
ones.  All these diagrams are proportional to the squared longitudinal form
factors of both colliding nuclei~\cite{CKTr}. Following the estimations 
presented in reference~\cite{CKTr}, the RHIC energies are just in the order of 
magnitude needed to observe this effect. 

Unfortunately, all estimations are model dependent, since the numerical weight
of the contribution of the multipomeron diagrams remains rather unclear due to
the many unknown vertices in these diagrams. However, the number of unknown
parameters can be reduced in some models, and, for example, in reference~\cite{CKTr}
the Schwimmer model~\cite{Schw} was used for the numerical calculations.
 


Another possibility to estimate the contribution of the
diagrams with Pomeron interaction comes~\cite{JUR,JUR1,BP,JDDSh,BJP}
from Percolation Theory. The percolation approach and its previous version, 
the String Fusion Model~\cite{SFM,SFM1,SFM2}, predicted the multiplicity
suppression seen at RHIC energies long before any RHIC data were taken.

New calculations of inclusive densities and multiplicities in percolation 
theory, both in $pp$~\cite{CP1,CP2} and in heavy ion collisions~\cite{CP2,CP3},
are in a good agreement with the experimental data for a wide energy region.


In the percolation approach one assumes that if two or several Pomerons 
overlap in transverse space, they fuse in only one Pomeron. When all 
quark-gluon strings (cut Pomerons) are overlapping, the inclusive density 
saturates, reaching its maximal value at a given impact parameter.

This approach has only one free parameter, $\eta$ ~\cite{JUR,JUR1,BP,JDDSh,BJP}:
\begin{equation}
\eta = N_s\cdot\frac{r^2_s}{R^2}\cdot\langle r(y) \rangle \;,
\end{equation}
with $N_s$ the number of produced strings, $r_s$ the string transverse 
radius, and $R$ the radius of the overlapping area. The factor 
$\langle r(y) \rangle$ accounts for the fact that the parton density near 
the ends of the string is smaller than in the central region, where we 
fix $r(0) = 1$. At large rapidities, we have $N_s$ strings with different
parton densities, $r_i(y)$, and
\begin{equation}
N_s\cdot\langle r(y) \rangle = \sum_{i=1}^{N_s} r_i(y) \;.
\end{equation}

As a result, the bare inclusive density $dn/dy \vert_{bare}$ gets reduced,
and we obtain:
\begin{equation}
 dn/dy = F(\eta) \cdot dn/dy \vert_{bare} \; ,
\end{equation}
with \cite{BP1}
\begin{equation}
 F(\eta) = \sqrt{\frac{1 - e^{-\eta}}{\eta}} \; .
\end{equation}

The number of quark-gluon strings (cut pomerons) increases with the centrality 
of the nucleus-nucleus collisions. If the secondaries are independently produced
in every quark-gluon string, then the ratio of yields of different particles should
not depend on the centrality of the collision. At not very high energies, as 
quark-gluon strings are usually not overlapped (see Fig.~3a), the parameter 
$\eta$ is rather smaller. At very high energies, on the contrary, the value
of $\eta$ is large, and in every heavy ion collision quark-gluon strings can be
overlapped, as it is shown in  Fig.~3b.

However, experimentally, even at not very high energies (e.g. see~\cite{HUA}),
the yield of $\overline{\Xi}^+$ is more strongly enhanced than that of 
$\overline{\Lambda}$, and also the yield of $\Omega^- + \overline{\Omega}^+$
shows a stronger enhancement than the yield of $\overline{\Xi}^+$. 
This means that one additional source of multistrange hyperons must be at work
from collective interactions of several strings.

At intermediate energies, we can think of rare situations when a few 
strings overlap in a small volume (see Fig.~3c). This can be possibly considered
as a QGP state effect. This can occur when the density of produced quark-gluon
strings is several times larger than its average value. 
\begin{figure}[htb]
\centering
\vskip -3.5cm
\includegraphics[width=.42\hsize]{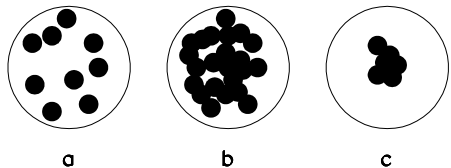}
\caption{\footnotesize
(a) Average heavy ion collisions at intermediate energies;
(b) Average heavy ion collisions at very high energies;
(c) Rare configuration when several strings overlap in a small volume.
Quark-gluon strings are shown by black points.}
\end{figure}

Contrary to what happens in the case shown in Fig.~3a, for the configurations
in Figs.~3b and~3c we can assume that the quark-gluon plasma has been produced.
In the next section we will show that these rare 
configurations can possibly give a large contribution in the cases of $\overline{\Xi}^+$ 
and, especially $\overline{\Omega}^+$, production.

When we consider central Pb+Pb collisions, we see that the experimental data on the
ratios in Eqs. (1) and (2) are in disagreement with the standard QGSM predictions
described in Section 2. 
The experimental ratios show very strong energy 
dependences as a function of centrality in the CERN SpS-RHIC energy interval.
To account for this effect, we can assume that some new contribution for
multistrange antihyperon formation  appears in this case. For example, the
three strange antiquarks needed for the formation of $\overline{\Omega}^+$
could be taken at some energies out of three different quark-gluon strings,
and not from the fragmentation of the same string. This could be possible if
the multiplicity of newly produced strange-antistrange pairs would decreases
with the number of pairs faster than, say, a Poissonian distribution.
This new contribution is not included in the standard QGSM scheme, and
it can be identified as a QGP formation signature. 

Probably, when the energy increases the distribution of the number 
of strange-antistrange pairs corresponding to this new contribution
decreases not so fast, and so the contribution from one quark-gluon string
will dominate over the collective contributions. However, this effect appears
to be very model dependent, so we will not consider its details in this paper. 

In order to account for the percolation effects in the QGSM, it is 
technically more simple~\cite{MPS1} to consider in the central 
region the maximal number of Pomerons $n_{max}$ emitted by one nucleon
that can be cut. These cut Pomerons lead then to the different final 
states, and the contributions of all diagrams with $n \leq n_{max}$ are 
accounted for as at lower energies. Larger number of Pomerons 
$n > n_{max}$ can also be emitted obeying the unitarity constraint, but due to 
fusion in the final state (at the quark-gluon string stage), the cut of 
$n > n_{max}$ Pomerons results in the same final state as the cut of 
$n_{max}$ Pomerons.
 
The QGSM fragmentation formalism allows one to 
calculate the integrated over $p_T$ spectra of different secondaries as
function of the rapidity. The number of strings that can be used for the secondary 
production should increase with the initial energy, as it was shown 
in~\cite{JDDCP}.


The contribution coming from the coherent interaction of two nuclear clusters  
\cite{Efre} has been estimated following~\cite{???} to be not larger than (20-30)\%.

\section{Theoretical Results and Selection Trigger for Quark-Gluon Plasma 
Formation Events}

\subsection{NA49 Collaboration Data}

The NA49 Collaboration obtained experimental data \cite{NA49a,NA49b} for
yields of $\overline{\Lambda}$ and $\overline{\Xi}^+$
hyperons in midrapidity region ($\vert y \vert < 0.4$ for  
$\overline{\Lambda}$, and $\vert y \vert < 0.5$ for  
$\overline{\Xi}^+$) in central C+C, Si+Si, and Pb+Pb collisions 
(5\% centrality for $\overline{\Lambda}$, and 10\% centrality for 
$\overline{\Xi}^+$) at 158 GeV per nucleon. These results 
are presented in Table~1, together with the QGSM results calculated
for the same rapidities and centralities.
 
We see a reasonable agreement for
secondary  $\overline{\Lambda}$, but the calculated yields of  
 $\overline{\Xi}^+$ are significantly smaller than those measured
by the NA49 Collaboration. This disagreement can have very important implications
that we will discuss in more detail in the next subsection.


\begin{center}
\vskip 5pt
\begin{tabular}{|c||c|c|c|} \hline
Collision & QGSM & NA49 Collaboration \\ \hline

C+C $\to \overline{\Lambda}$ & 0.064 & 
$0.064 \pm 0.003 \pm 0.010$ \\ \hline

Si+Si $\to \overline{\Lambda}$ & 0.17 &
$0.16 \pm 0.007 \pm 0.038$ \\ \hline

Pb+Pb $\to \overline{\Lambda}$ & 2.05 &
$1.4 \pm 0.3 \pm 0.2$ \\ \hline

Pb+Pb $\to \overline{\Xi}^+$ & 0.16 &
$0.31 \pm 0.03 \pm 0.03$  \\ \hline

\hline
\end{tabular}
\end{center}
Table 1: Experimental data~\cite{NA49a,NA49b} by the NA49 Collaboration
for $\overline{\Lambda}$ and $\overline{\Xi}^+$ production in central
(5\% centrality) C+C, Si+Si, and Pb+Pb collisions at 158~GeV per nucleon,
and the corresponding QGSM results.

\subsection{NA57 Collaboration Data}

The NA57 Collaboration obtained the experimental data~\cite{NA57} for strange
and multistrange hyperon $\overline{\Lambda}$, $\overline{\Xi}^+$, and
$\overline{\Omega}^+$ yields in the midrapidity region $\vert y \vert < 0.5$
in minimum bias $p$+Be and $p$+Pb reactions, and in central (5\% centrality)
Pb+Pb collisions, at 158 GeV per nucleon.

These data are presented in Table 2, together with the corresponding
QGSM calculations.

Unfortunately, the data obtained by NA49 and NA57 Collaborations are not
compatible, probably due to different experimental event selection. As one can
see in Tables 1 and 2, the values of $dn/dy$ for the different hyperons obtained
by one collaboration are far outside the error bars of the values obtained at the
same centrality by the other collaboration.


\vskip 5pt

\begin{center}
\vskip 5pt
\begin{tabular}{|c||c|c|c|} \hline
Collision & QGSM &  NA57 Collaboration\\ \hline

p+Be $\to \overline{\Lambda}$ & 0.010 & 
$0.011 \pm 0.0002 \pm 0.0001$ \\ \hline

p+Be $\to \overline{\Xi}^+$ & 0.00081 &
$0.00077 \pm 0.0001 \pm 0.0001$ \\ \hline

p+Be $\to \overline{\Omega}^+$ & 0,000042 &
$0,00074 \pm 0.0002 \pm 0.001$ \\ \hline

p+Pb $\to \overline{\Lambda}$ & 0.019 & 
$0.015 \pm 0.001 \pm 0.002$ \\ \hline

p+Pb $\to \overline{\Xi}^+$ & 0.0015 &
$0.0012 \pm 0.001 \pm 0.001$ \\ \hline

p+Pb $\to \overline{\Omega}^+$ & 0,000076 &
$0,0095 \pm 0.003 \pm 0.001$ \\ \hline

Pb+Pb $\to \overline{\Lambda}$ & 2.05 & 
$2.44 \pm 0.14 \pm 0.24$ \\ \hline

Pb+Pb $\to \overline{\Xi}^+$ & 0.16 &
$0.51 \pm 0.04 \pm 0.05$ \\ \hline

Pb+Pb $\to \overline{\Omega}^+$ & 0,006668 &
$0,16 \pm 0.04 \pm 0.02$ \\ \hline

\hline
\end{tabular}
\end{center}
Table 2: Experimental data~\cite{NA57} obtained by the NA57 Collaboration
for $\overline{\Lambda}$, $\overline{\Xi}^+$, and $\overline{\Omega}^+$
production in $p$+Be, $p$+Pb, and in central (5\% centrality) Pb+Pb collisions
at 158~GeV per nucleon, and the corresponding QGSM results.

Here again one can see that the calculated yields of $\overline{\Lambda}$ are in
agreement with the experimental data on the level of 20-30\% accuracy. Inclusive
densities of $\overline{\Xi}^+$ hyperons are reasonably reproduced for the cases
of p+Be and p+Pb collisions, but are several times underestimated in the case of
central Pb+Pb interactions. For $\overline{\Omega}^+$ production in central 
Pb+Pb collisions, the disagreement is larger than one order of magnitude.
The reason to discuss the production of antihyperons in these processes
is that since a large baryon charge already exists in the initial state, some hyperons
can be produced via the final state interaction of any baryon with a strange
meson, but, on the contrary, the production of multistrange antihyperon production
would be a signal of the production of several strange antiquarks in some small
space-time volume.

These effects are more evidently seen in Fig.~4, where the experimental ratios of 
$\overline{\Omega}^+ / \overline{\Lambda}$ and $\overline{\Xi}^+ / \overline{\Lambda}$
as functions of the number of participating nucleons, $N_w$, together
with the corresponding QGSM results, shown by solid curves. The first two left points
in every panel correspond to p+Be and p+Pb collisions, and the other points correspond
to Pb+Pb interactions with different centralities.
\begin{figure}[htb]
\centering
\includegraphics[width=.49\hsize]{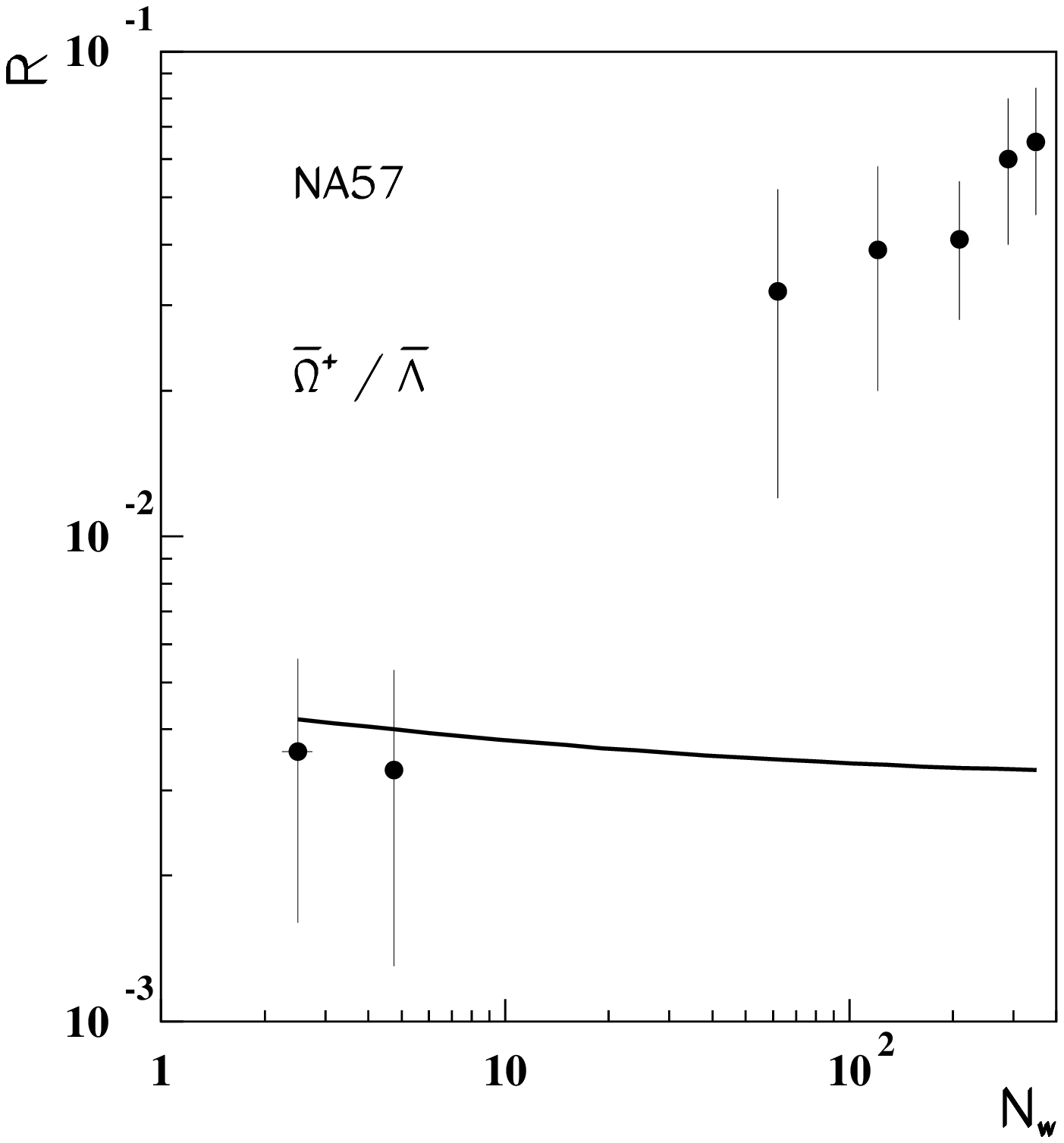}
\includegraphics[width=.49\hsize]{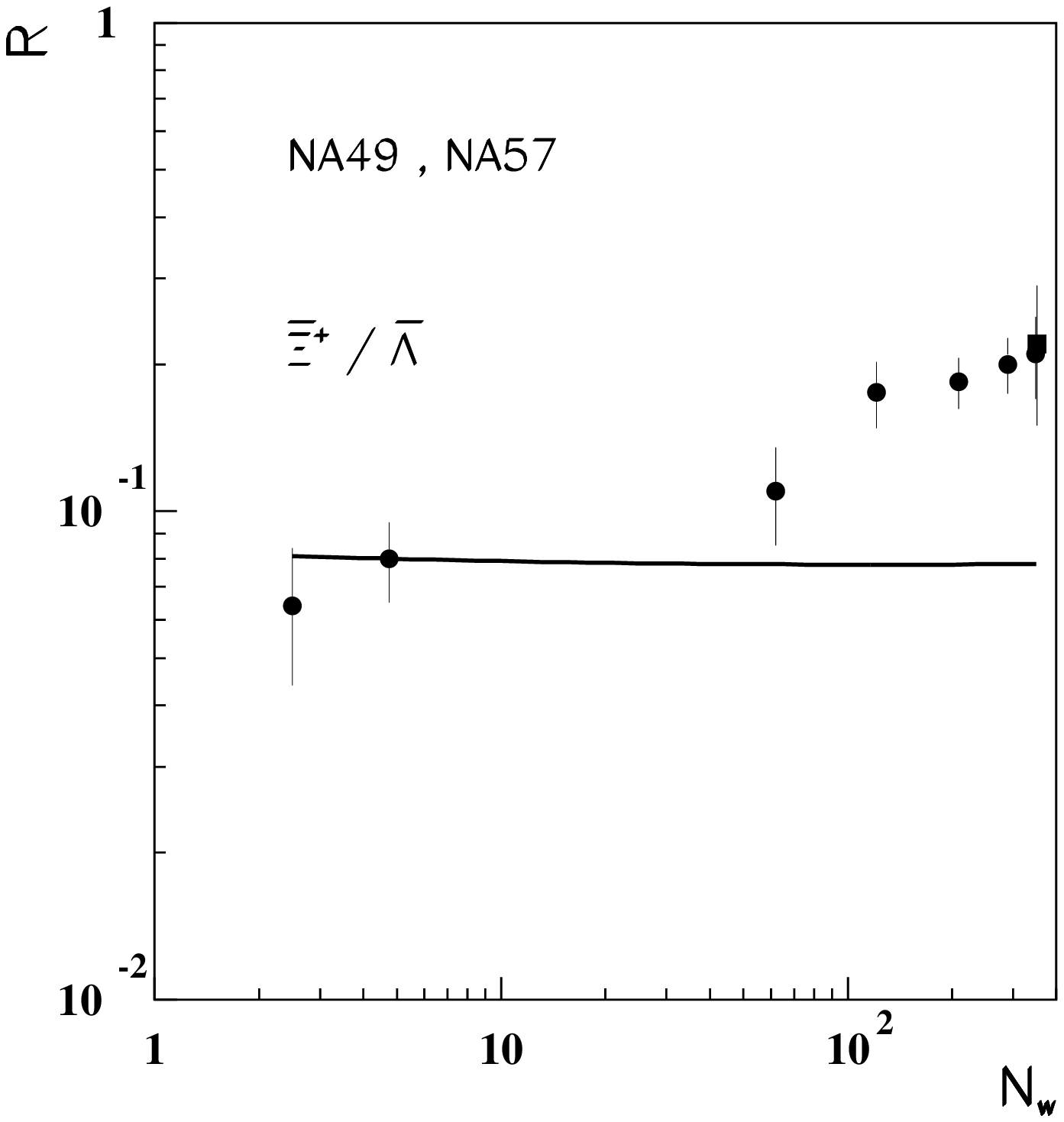}
\vskip -.3cm
\caption{\footnotesize
Ratios of $\overline{\Omega}^+$ to $\overline{\Lambda}$ (left panel), and
of $\overline{\Xi}^+$ to $\overline{\Lambda}$ (right panel) as functions of the
number of wounded nucleons, $N_w$. The experimental data for p+Be, p+Pb, and
for Pb+Pb with different centralities measured by the NA57 Collaboration (points)
and by the NA49 Collaboration (squares) are presented, together with the
corresponding QGSM results, shown by solid curves.}
\end{figure}

We consider the disagreement of the QGSM results with the experimantal values
as a signal of the quantitatively large new contribution of $\overline{\Omega}^+$ and
$\overline{\Xi}^+$ production to the inclusive cross section. This new contribution
could appear when strange antiquarks are taken from different Pomerons (quark-gluon strings), 
in some collective interaction (see Fig.~3c) that can be seen as at the origin 
of Quark-Gluon Plasma formation.

Thus, the selection of events with $\overline{\Xi}^+$, and especially with 
$\overline{\Omega}^+$ production, would allow the definition of a sample
enriched by Quark-Gluon Plasma formation events, and such a sample could be compared
to a sample of events where only $\overline{\Lambda}$ hyperons are produced, to unravel
QGP definite features.

\subsection{STAR Collaboration Data}

Hyperon production at higher energies in midrapidity region was measured at 
RHIC. The data~\cite{STAR} obtained by the STAR Collaboration for Au+Au collisions
at $\sqrt{s_{NN}}$ = 62.4 GeV are presented in Fig.~5.
\begin{figure}[htb]
\centering
\includegraphics[width=.49\hsize]{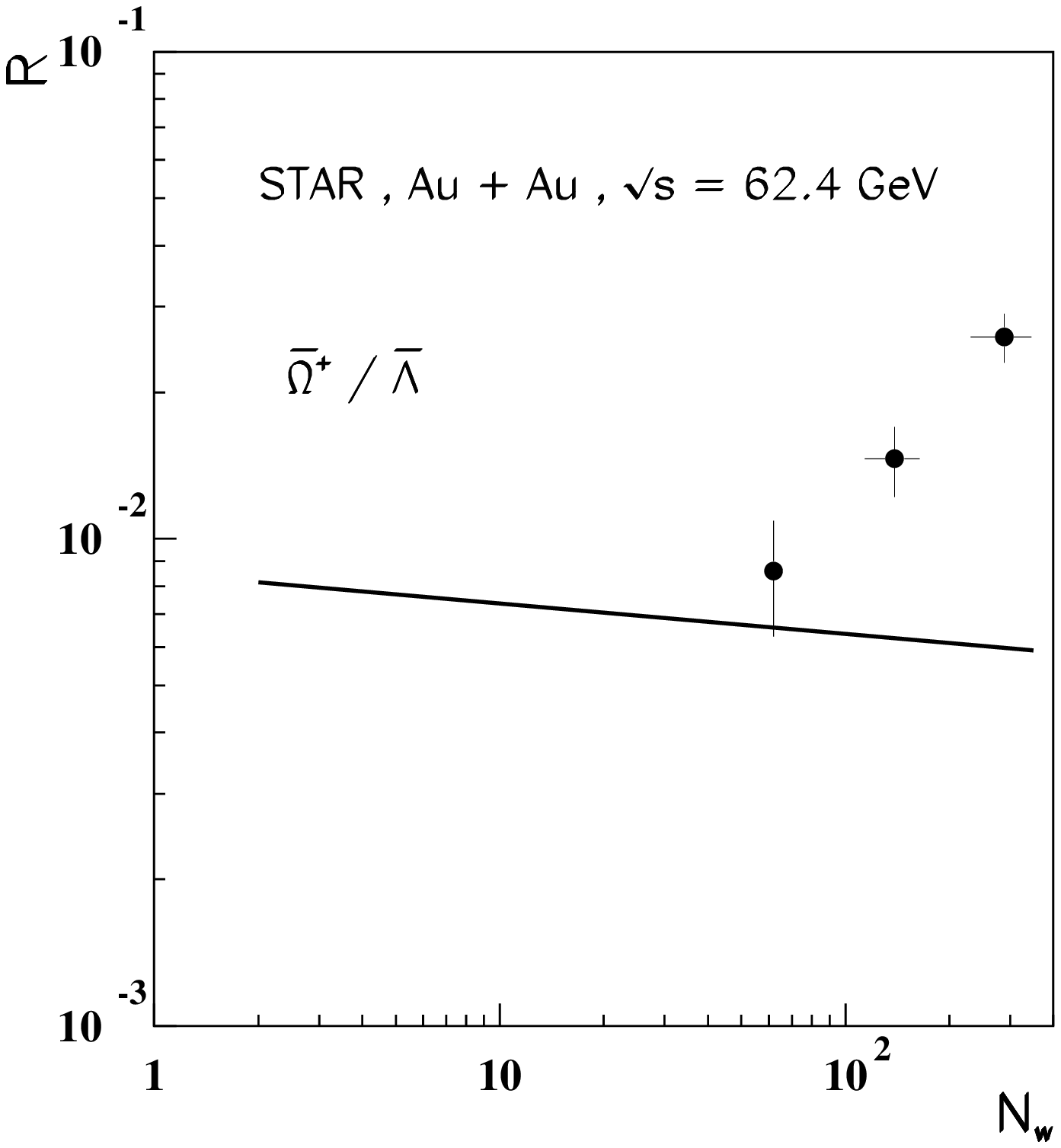}
\includegraphics[width=.49\hsize]{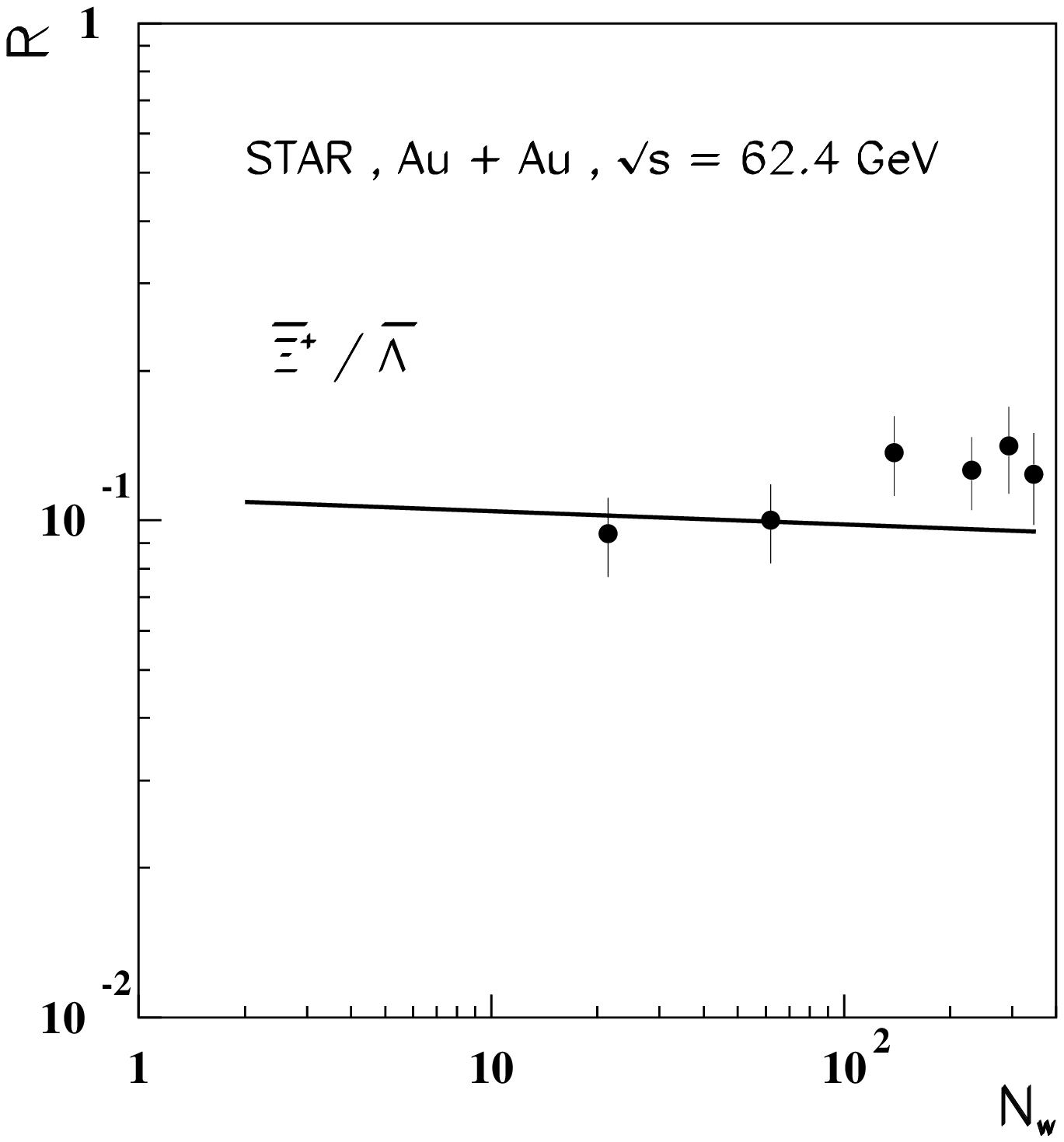}
\vskip -.3cm
\caption{\footnotesize
The experimental points obtained by the STAR Collaboration on the ratios of
$\overline{\Omega}^+$ to $\overline{\Lambda}$ (left panel), and of
$\overline{\Xi}^+$ to $\overline{\Lambda}$ (right panel), as functions of the
number of wounded nucleons, $N_w$, for Au+Au collisions with different centralities,
together with the corresponding QGSM results, shown by solid curves.}
\end{figure}

Now the experimental ratios of $\overline{\Omega}^+$ to $\overline{\Lambda}$ remain larger
than the QGSM predictions at high centralities, and, so, the new mechanism by which
the three strange antiquarks needed for $\overline{\Omega}^+$ formation are
taken from different Pomerons is still important, though its relative contribution 
decreases in comparison with CERN SpS energies.

On the other hand, the experimental ratios of $\overline{\Xi}^+$ to $\overline{\Lambda}$
are now in agreement with the QGSM model calculations, probably meaning that the
production of $\overline{\Xi}^+$ inside one only Pomeron becomes more effective
at this energy than the possibility of taking two strange antiquarks from different
Pomerons.

This trend is confirmed by the STAR Collaboration data~\cite{STAR1} at $\sqrt{s_{NN}}$ 
= 200 GeV, that are presented in Table~3 and in Fig.~6.
\begin{center}
\vskip 5pt
\begin{tabular}{|c||c|c|c|} \hline
Collision & QGSM &  STAR Collaboration\\ \hline

Cu+Cu $\to \overline{\Lambda}$ & 3.34 & 
$3.79 \pm 0.37 $ \\ \hline

Cu+Cu $\to \overline{\Xi}^+$ & 0.41 &
$0.52 \pm 0.8 $  \\ \hline

Au+Au $\to \overline{\Lambda}$ & 13.3 &
$14.7 \pm 0.9 $ \\ \hline

Au+Au $\to \overline{\Xi}^+$ & 1.66 &
$  - $  \\ \hline

\hline
\end{tabular}
\end{center}
Table 3: Experimental data~\cite{STAR1} by the STAR Collaboration on
$\overline{\Lambda}$ and $\overline{\Xi}^+$ production in central Cu+Cu and
Au +Au collisions at $\sqrt{s_{NN}}$ = 200~GeV, together with the corresponding
QGSM results.

The ratios of $\overline{\Xi}^+$ to $\overline{\Lambda}$  at  $\sqrt{s_{NN}}$ = 
200 GeV are in agreement with  the QGSM calculations.
\begin{figure}[htb]
\centering
\includegraphics[width=.49\hsize]{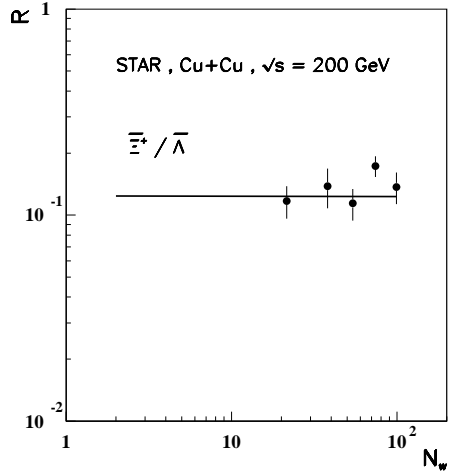}
\vskip -.3cm
\caption{\footnotesize
The experimental points obtained by the STAR Collaboration on the ratios
of $\overline{\Xi}^+$ to $\overline{\Lambda}$ as functions of the
number of wounded nucleons, $N_w$, for Au+Au collisions with different centralities,
together with the corresponding QGSM results, shown by solid curve.}
\end{figure}

\section{Predictions for LHC and Conclusion}

In nucleus-nucleus collisions, the number of quark-gluon strings (cut pomerons) increases
with centrality. If the secondaries are independently produced in each quark-gluon string,
the ratio of yields of different particles should not depend on centrality. However, it is
experimentally clear (e.g. see~\cite{HUA}, and Figs.~4 and 6 ) that the yield of
$\overline{\Xi}^+$ is more strongly enhanced than that of 
$\overline{\Lambda}$, the same as the yield of $\Omega^- + \overline{\Omega}^+$ is more enhanced
than that of $\overline{\Xi}^+$. This shows that an additional source of 
multistrange hyperons originated by the collective interactions of several strings must also
play an actile role, and then this additional source of multistrange hyperons can possibly
be considered as a QGP signature.

If this scenario in which collective effects among different quark-gluon strings help in
explaining the energy dependence from CERN SpS to RHIC of the production ratios of
$\overline{\Omega}^+$ to $\overline{\Lambda}$ and of $\overline{\Xi}^+$ to $\overline{\Lambda}$, would
be confirmed, one could not expect a significant dependence of these ratios on centrality
(see Fig.~4 and left panel of Fig.~5), neither significant differences of these 
ratios at LHC energies from the QGSM predictions, the expected values being 
$R(\overline{\Omega}^+/\overline{\Lambda}) \sim 0.14$ and 
$R(\overline{\Xi}^+/\overline{\Lambda}) \sim 0.014$.

A situation similar to the one shown in Figs.~4 can appear at LHC in Pb+Pb collisions
in the midrapidity region for charmed antibaryon production, that is, 
for the ratios  $\overline{\Omega}^+_c = \overline{c}\overline{s}\overline{s}$ to 
$\overline{\Lambda}_c $, and $\overline{\Xi}^+_c$ to $\overline{\Lambda}_c$.

In summary, we can assume that the considered Quark-Gluon Plasma formation events
are rather rare, but at the same time they can be clearly identified by a well-defined
selection trigger. At CERN SpS energy, such a trigger
can be the $\overline{\Omega}^+$ production in central Au+Au collisions, where
a strong centrality dependence of the ratio $\overline{\Omega}^+$ to $\overline{\Lambda}$
is apparent (see Fig.~4a).

{\bf Acknowledgements}

We are grateful to C. Pajares for useful discussions and
comments and to N.I. Novikova for technical help.
This paper was supported by Ministerio de Econom\'i a y
Competitividad of Spain (FPA2011$-$22776), the Spanish
Consolider-Ingenio 2010 Programme CPAN (CSD2007-00042),
by Xunta de Galicia, Spain (2011/PC043), by the State
Committee of Science of the Republic of Armenia
(Grant-13-1C023), and, partially, by grant RSGSS-3628.2008.2.



\begin{thebibliography}{**}

\bibitem{KTM} A.B. Kaidalov and K.A. Ter-Martirosyan, Yad. Fiz. {\bf 39},
1545 (1984); {\bf 40}, 211 (1984).

\bibitem{KaPi} A.B. Kaidalov and O.I.~Piskounova, Yad. Fiz. {\bf 41}, 1278
(1985); Z. Phys. {\bf C30},145 (1986).

\bibitem{Sh} Yu.M. Shabelski, Yad. Fiz. {\bf 44}, 186 (1986).

\bibitem{AMPS} G.H. Arakelyan, C. Merino, C. Pajares, and Yu.M.~Shabelski,
Eur. Phys. J. {\bf C54}, 577 (2008) and hep-ph/0709.3174.

\bibitem{MPS} C. Merino, C. Pajares and Yu.M.~Shabelski, Eur. Phys. J.
{\bf C71}, 1652 (2011).

\bibitem{KTMS} A.B. Kaidalov, K.A. Ter-Martirosyan, and Yu.M.~Shabelski,
Yad. Fiz. {\bf 43}, 1282 (1986).

\bibitem{Sh1} Yu.M. Shabelski, Z. Phys. {\bf C38}, 569 (1988).

\bibitem{Sha}
 Yu.M. Shabelski, Yad. Fiz. {\bf 50}, 239 (1989).

\bibitem{Shab} Yu.M. Shabelski, Z. Phys. {\bf C57}, 409 (1993).

\bibitem{JDDS} J. Dias de Deus and Yu.M. Shabelski, Yad. Fiz. {\bf 71}, 191
(2008).

\bibitem{???}  G.H. Arakelyan, C. Merino, and Yu.M.~Shabelski,
arXiv:1305.0388 [hep-ph].


\bibitem{BSh} V.M. Braun and Yu.M.~Shabelski,
Int. J. Mod. Phys. {\bf A3}, 2117 (1988).

\bibitem{Alk} G.D. Alkhazov {\it et al.}, Nucl. Phys. {\bf A280}, 365 (1977).

\bibitem{CKTr} A. Capella, A. Kaidalov, and J. Tran Thanh Van, Heavy Ion
Phys. {\bf 9}, 169 (1999).


\bibitem{HUA} Huan Z. Huang, J. Phys. {\bf G30}, 401 (2004).

\bibitem{AGK} V.A. Abramovsky,  V.N. Gribov, and O.V.~Kancheli, Yad. Fiz.
{\bf 18}, 595 (1973).

\bibitem{Kai} A.B. Kaidalov, Sov. J. Nucl. Phys. {\bf 45}, 902 (1987);
Yad. Fiz. {\bf 43}, 1282 (1986).

\bibitem{ACKS} G.H. Arakelyan, A. Capella, A.B.~Kaidalov, and
Yu.M.~Shabelski, \newline
Eur. Phys. J. C~{\bf26}, 81 (2002) and hep-ph/0103337.

\bibitem{Sh3} Yu.M. Shabelski, Yad.Fiz. {\bf 26}, 1084 (1977);
Nucl. Phys. {\bf B132}, 491 (1978).

\bibitem{BT} L. Bertocchi and D. Treleani, J. Phys. {\bf G3}, 147 (1977).

\bibitem{Weis} J. Weis, Acta Phys. Polonica {\bf B7}, 85 (1977).

\bibitem{Jar} T. Jaroszewicz {\it et al.}, Z. Phys. {\bf C1}, 181 (1979).

\bibitem{Phob} B.B. Back {\it et al.} (PHOBOS Collaboration), Phys. Rev. Lett. 
{\bf 85}, 3100 (2000).

\bibitem{Phen} K. Adcox {\it et al.} (PHENIX Collaboration), Phys. Rev. Lett. 
{\bf 86}, 500 (2001).

\bibitem{CMT} A. Capella, C. Merino, and J. Tran Thanh Van, Phys. Lett.
{\bf B265} (1991) 415.

\bibitem{Sh6} Yu.M. Shabelski, Z. Phys. {\bf C57}, 409 (1993).

\bibitem{AP} N. Armesto and C. Pajares, Int. J. Mod. Phys.
{\bf A15}, 2019 (2000).

\bibitem{Kan} O.V. Kancheli, JETP Lett. {\bf 18}, 274 (1973).

\bibitem{NNN} G.V. Davidenko and N.N. Nikolaev, Yad. Fiz. {\bf 24}, 772 (1976).

\bibitem{Schw} A. Schwimmer, Nucl. Phys. {\bf B94}, 445 (1975).

\bibitem{JUR} J. Dias de Deus, R. Ugoccioni, and A. Rodrigues, Phys. Lett. 
{\bf B458}, 402 (1999).

\bibitem{JUR1} J. Dias de Deus, R. Ugoccioni, and A. Rodrigues, Eur. Phys. J. 
{\bf C16}, 537 (2000).

\bibitem{BP} M.A. Braun and C. Pajares, Phys. Rev. Lett. {\bf 85}, 4864 (2000).

\bibitem{JDDSh} J. Dias de Deus and Yu.M. Shabelski, Eur. Phys. J. {\bf A20}, 
457 (2004).

\bibitem{BJP} P. Brogueira, J. Dias de Deus, and C. Pajares, Phys. Rev. 
{\bf C75}, 054908 (2007).

\bibitem{SFM} C. Merino, C. Pajares, and J. Ranft, Phys. Lett. {\bf B276}, 168 
(1992).

\bibitem{SFM1} H.J. M{\"o}hring, J. Ranft, C. Merino, and C. Pajares, 
Phys. Rev. {\bf D47}, 4142 (1993).

\bibitem{SFM2} N.S. Amelin, M.A. Braun, and C. Pajares, Z. Phys. {\bf C63}, 
507 (1994).

\bibitem{CP1} I. Bautista, C. Pajares, and J. Dias de Deus, Nucl. Phys. 
{\bf A882}, 44 (2012).

\bibitem{CP2} I. Bautista, J. Dias de Deus, G. Milhano, and C. Pajares, 
Phys. Lett. {\bf B715}, 230 (2012).

\bibitem{CP3} I. Bautista, C. Pajares, G. Milhano, and  J. Dias de Deus,
Phys. Rev. {\bf C86}, 034909 (2012).

\bibitem{BP1} M.A. Braun and C. Pajares, Eur. Phys. J. {\bf C16}, 2019 (2000).

\bibitem{MPS1} C. Merino, C. Pajares, and Yu.M.~Shabelski, Eur. Phys. J.
{\bf C59}, 691 (2009) and arXiv:0802.2195[hep-ph].

\bibitem{JDDCP} J. Dias de Deus and C. Pajares, Phys. Lett.
{\bf B695}, 211 (2012) and arXiv:1011.1099[hep-ph].

\bibitem{Efre} A.V. Efremov {\it et al.}, Phys. Atom. Nucl. {\bf 57}, 874 (1994).

\bibitem{NA49a}  T. Anticic {\it et al.} (NA44 Collaboration), Phys. Rev. {\bf C80},
034906 (2009) and arXiv:0906.0469[nucl-ex].

\bibitem{NA49b}  T. Anticic {\it et al.} (NA44 Collaboration), Phys. Rev. {\bf C78},
034918 (2008) and arXiv:0804.3770[nucl-ex].

\bibitem{NA57} F. Antinori {\it et al.} (NA57 Collaboration), J. Phys. {\bf G32},
427 (2006) and arXiv:0601021[nucl-ex].

\bibitem{STAR} M.M. Aggarwal {\it et al.} (STAR Collaboration), Phys. Rev.
{\bf C83}, 024901 (2011) and arXiv:1010.0146[nucl-ex].

\bibitem{STAR1} G. Agakishiev {\it et al.} (STAR Collaboration), Phys. Rev. Lett.
{\bf 108}, 072301 (2012) and arXiv:1107.2955[nucl-ex].

\end{thebibliography}
\end{document}